\documentclass[a4paper,10pt,twocolumn]{article}
\usepackage{epsf,amsmath}
\usepackage{amssymb}
\usepackage{indentfirst}
\makeatletter
\def\@biblabel#1{#1.}
\makeatother

\begin{document}

\title{\Large \bf Long Gamma-Ray Bursts and the Morphology of their Host Galaxies }

\author{A. I. Bogomazov$^{1}$\footnote{bogomazov@sai.msu.ru}, V. M. Lipunov$^{1}$\footnote{lipunov@sai.msu.ru}, A. V. Tutukov$^{2}$\footnote{atutukov@inasan.rssi.ru} \\
\small\it $^1$ Lomonosov Moscow State University, Sternberg
Astronomical Institute, \\ \small 119992, Universitetskij prospect, 13, Moscow, Russia \\
\small\it $^2$ Institute of Astronomy, Russian Academy of Sciences, \\ \small 119017, Pyatnitskaya ulitsa, 48, Moscow, Russia \\
}

\date{\begin{minipage}{15.5cm} \small
We present the results of population syntheses for binary stars
carried out using the ``Scenario Machine'' code with the aim of
analyzing events that may result in long gamma-ray bursts. We show
that the observed distribution of morphological types of the host
galaxies of long gamma-ray bursts can be explained in a model in
which long gamma-ray bursts result from the core collapse of
massive Wolf-Rayet stars in close binaries. The dependence of the
burst rate on galaxy type is associated with an increase in the
rate of stellar-wind mass-loss with increasing stellar
metallicity. The separation of binary components at the end of
their evolution increases with the stellar-wind rate, resulting in
a reduction of the number of binaries that produce gamma-bursts.
\end{minipage} } \maketitle \rm

\section{Introduction}

Although cosmic gamma-ray bursts (GRBs) have been studied for four
decades, the origin of this phenomenon is far from fully
understood. At present, according to observational limitations,
gamma-ray bursts are subdivided into short GRBs (durations less
than $\approx 2$ s) and long GRBs (durations more than $\approx 2$
s). This likely reflects different natures of the precursors of
the two types of GRBs \cite{kouveliotou1993a}.

Long GRBs, are probably associated with the core collapse of
massive Wolf-Rayet stars in close binaries, accompanied by the
formation of accretion disks around black holes
\cite{woosley1993a}. These events may be associated with the
formation of spinars\footnote{A spinar is a collapsed,
quasi-equilibrium object whose equilibrium is supported by the
balance of the centrifugal and gravitational forces. The evolution
of a spinar is determined by its magnetic field.}
\cite{lipunov2007a,lipunov2007b}. There exist two possible reasons
for fast rotation of the presupernova core: acceleration of the
rotation of the core of an evolving star if its angular momentum
is conserved, or the presence of a close companion to the
presupernova -- a Wolf-Rayet star\footnote{Cyg X-3 may serve as an
example of such a system.}. The second option seems more
appropriate, since the rotation of the cores of massive stars most
likely decelerates in the course of evolution
\cite{tutukov1969a,maeder2000a}, while binary orbital motion
enables fast rotation of the presupernova due to to tidal
synchronization of its rotation \cite{tutukov2003a,tutukov2004a}.

Below we will consider Kerr black holes, which we define as
objects with effective Kerr parameters $a$ satisfying the
condition

\begin{equation}
\label{kerr} a=\frac{I\Omega}{GM_{BH}^2/c}\ge 1,
\end{equation}

\noindent where $I$ is the black hole's moment of inertia,
$\Omega$ the angular frequency of its rotation, $M_{BH}$ its mass,
$G$ the gravitational constant, and $c$ the velocity of light.
This condition for the Kerr parameter is necessary for the
formation of a massive accretion disk or a spinar, if a strong
enough magnetic field is present. It is clear that (\ref{kerr})
does not describe any real black hole, but provides a
characteristic of the angular momentum of both the non-degenerate
presupernova and the nascent black hole. In the absence of
accurate estimates of the masses of collapsing cores and the
fractions of the stellar mass falling inside the event horizon
during the formation of black holes, we assume that the orbital
period of a binary prior to the supernova explosion must be less
than $\approx 1-3$ day \cite{tutukov2003a,tutukov2004a}.

Sokolov et al. \cite{sokolov2001a} published observations of host
galaxies of long GRBs obtained with the 6-m telescope of the
Special Astrophysical Observatory of the Russian Academy of
Sciences. They discovered that long GRBs occur in galaxies with
very intense star formation and significant internal absorption,
and with luminosities not exceeding the luminosity of the Milky
Way.

Observational data on host galaxies of long GRBs and Type Ic
supernovae (SN Ic) obtained with the Hubble Space Telescope are
provided in \cite{fruchter2006a}. It was assumed that SN Ic and
long GRBs should occur in similar environments. However, the
observational data show that this is not the case. First, core
collapse supernovae (such as SN Ic) are located in blue regions of
their host galaxies, while long GRBs are located in the brightest
regions of the galaxies. Second, most long GRBs occur in dwarf
irregular galaxies, while SN Ic happen in both dwarf irregulars
and giant spirals \cite[Fig. 1]{fruchter2006a}. Only one of 42
long GRBs occurred in a giant spiral galaxy, with the others
occurring in dwarf irregular galaxies; if the redshift of the host
galaxies is restricted to 1.2, one in 18 long GRBs occurred in
giant spirals.

The Galactic rates of events resulting in rotational collapse
accompanied by the formation of rapidly spinning relativistic
objects were analyzed in \cite{bogomazov2007a}, where it was
concluded that these rates are sufficient to explain the cosmic
rate of observed GRBs. The aim of the present paper is to show
that the observed correlation between host-galaxy and the rate of
long GRBs can be explained in a model in which long GRBs result
from core collapse in Wolf-Rayet stars in close binaries.

It is known that stellarmetallicity has a very strong effect on
the stellar wind, and that the metallicity of galaxies grows with
their mass and age. One example of a population-synthesis study of
the effect of metallicity on stellar wind is \cite{vink2006a}, and
a recent empirical study of the metallicities of star-forming
galaxies can be found in \cite{vaduvescu2007a} (see also, e.g.,
\cite{kobulnicky2004a}).

In the first half of the 1990s, radio observations were made of
the pulsar J0045-7319 in the Small Magellanic Cloud, which is in a
close binary with a B1 main-sequence star \cite{kaspi1996a}. The
estimated mass-loss rate by the optical star does not exceed
$10^{-10}M_{\odot}/\mbox{yr}$, two orders of magnitude lower than
for stars in the Milky Way (see, e.g., \cite{tutukov2003b}).

For our analysis, we introduce a luminous-mass limit separating
low- and high-metallicity galaxies (in our model, galaxies with
``strong'' and ``weak'' stellar winds). This limit corresponds
approximately to the observational data of
\cite{vaduvescu2007a,kobulnicky2004a}. Following
\cite{schechter1976a} we assume a mass function for the galaxies
(see, e.g., also \cite{salucci1999a,bell2003a})

\begin{equation}
\label{galmasfunc} \frac{dN}{d \log M}\approx\mbox{const}, \quad
10^5 M_{\odot} \le M \le 10^{11} M_{\odot},
\end{equation}

\noindent where $M$ is the mass of luminous matter in the galaxy
and $N$ the number of galaxies.

In addition, we introduce a maximum mass for galaxies that can
host long GRBs. Since the most massive galaxies usually have the
highest metallicities, and hence the strongest stellar winds, the
two components may be too far apart at the time of the supernova
explosion to satisfy condition (\ref{kerr}). For obvious reasons
[see (\ref{galmasfunc})], the input to the total rate of events
calculated using this mass function will be the highest for the
most massive galaxies in the range given above. Therefore,
adopting an upper limit for the masses of galaxies that can host
long GRBs can substantially affect the results.

\section{Population synthesis}

\begin{figure}[h!]
\hspace{0cm} \epsfxsize=0.45\textwidth\centering \epsfbox{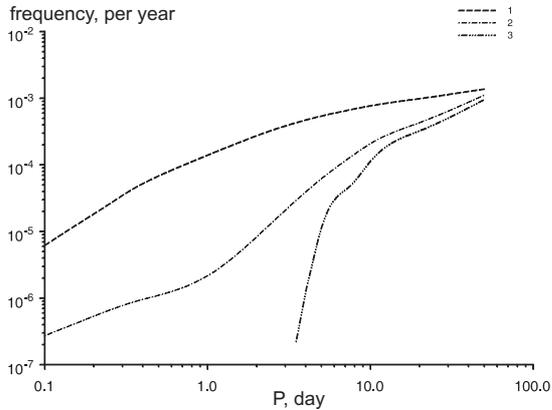}
\vspace{0cm}\caption{ Dependence of the Galactic rate of core
collapses of Wolf-Rayet stars in close binaries accompanied by the
formation of black holes with large orbital angular momenta on the
orbital period at the time of the supernova. Curve \emph{1} is
computed for stellar-wind model 1, curve \emph{2} for stellar-wind
model 2, and curve \emph{3} for stellar-wind model 3. }
\label{freq}
\end{figure}

We used the ``Scenario Machine'' for our population synthesis,
synthesizing $10^6$ binaries for each set of initial parameters.

Since the ``Scenario Machine'' has been described many times in
previous publications, we will only note here the most important
parameter for our study, which especially affects the results of
numerical modeling of the objects under investigation. A detailed
description of the ``Scenario Machine'' can be found in
\cite{Lipunov1996,Lipunov2007c}.

This critical parameter is the stellar wind, which strongly
influences the semi-major axis of the binary orbit. We used the
following stellar-wind models, which differ in the amount of
matter lost by a massive star in the course of its evolution:

\begin{itemize}

\item {\it Wind model 1.} This model corresponds to scenario A in
\cite{bogomazov2007a,Lipunov2007c}. The amount of matter lost via
the wind is small. The total fractional mass loss in the
main-sequence, supergiant, and Wolf-Rayet stages does not exceed
30\% of the initial stellar mass.

\item {\it Wind model 2.} This model corresponds approximately to
scenario C in \cite{bogomazov2007a,Lipunov2007c}, except that the
star loses 70\% of its envelope mass in each evolutionary stage.

\item {\it Wind model 3.} This model corresponds to scenario C in
\cite{bogomazov2007a,Lipunov2007c}. The star loses its entire
envelope in each evolutionary stage. This mean that, at the end of
the evolution, the total mass loss can be more than half the
initial mass of the star.

\end{itemize}

The assumed minimum\footnote{ If a star in the course of its
evolution accretes mass from a companion, such that its final mass
exceeds its initial mass, we consider the maximum mass of the star
after the completion of accretion. } initial mass of stars
producing black holes is $25 M_{\odot}$.

\section{Results}

\begin{figure}[h!]
\hspace{0cm} \epsfxsize=0.45\textwidth\centering \epsfbox{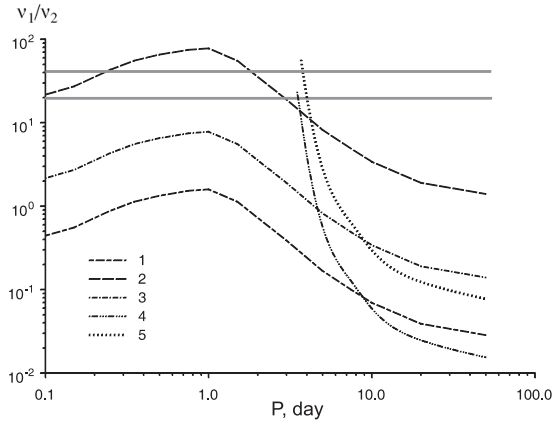}
\vspace{0cm}\caption{ Ratio of the rates of GRBs in galaxies of
different types $\nu_1/\nu_2$, calculated taking into account the
mass function of the galaxies [see (\ref{galmasfunc})]. $\nu_1$ is
the rate of GRBs in dwarf galaxies (with weak stellar wind), and
$\nu_2$ the rate of GRBs in giant galaxies (with strong stellar
wind). The horizontal line shows the observed ratio of the numbers
of long GRBs in dwarf irregulars and giant spirals (equal to 42 if
selection effects are not taken into account, and to 18 if the red
shift of the host galaxies is limited to z = 1.2). The numbers
label curves computed for various parameter values. Curve
\emph{1}: stellar-wind model 1 is used for the weak wind in dwarf
galaxies, and stellar-wind model 2 for the strong wind in giant
galaxies, with the limiting value of \emph{a} in the galaxy mass
function separating dwarf galaxies with weak wind and giant
galaxies with strong wind set to $10^9M_{\odot}$, and the upper
limit of the host-galaxy mass for GRBs $5\cdot 10^{10}M_{\odot}$.
Curve \emph{2}: stellar-wind model 2 was used for the strong wind
and stellar-wind model 1 for the weak wind, with $a=5\cdot
10^9M_{\odot}$, and $b=10^{10}M_{\odot}$. Curve \emph{3}:
stellar-wind model 2 was used for the strong wind and stellar-wind
model 1 for the weak wind, with $a=5\cdot 10^9M_{\odot}$ and
$b=5\cdot 10^{10}M_{\odot}$. Curve \emph{4}: stellar wind model 3
was used for the strong wind and stellar wind model 2 for the weak
wind, with $a=10^9M_{\odot}$ and $b=10^{11}M_{\odot}$. Curve 5:
stellar-wind model 3 was used for the strong wind and stellar-wind
model 2 for the weak wind, with $a=5\cdot 10^9M_{\odot}$ and
$b=10^{11}M_{\odot}$. } \label{rat}
\end{figure}

Figure 1 shows the dependence of the Galactic rate of core
collapses of Wolf-Rayet stars in close binaries accompanied by the
formation of black holes with large orbital angular momenta on the
orbital period at the time of the supernova. Curves {\it 1, 2, 3}
were computed for stellar-wind models 1, 2, 3, respectively. The
plot shows that this rate in close binaries (with orbital periods
$P<10$ day) varies very strongly with the mass outflow rate in the
course of the evolution of the normal star.

Figure 2 shows the ratio of the rates of GRBs in different types
of galaxies computed using the galaxy mass function [see
(\ref{galmasfunc})]. This plot shows that, within the existing
uncertainties in the observed relative rates of long GRBs and
mass-loss scenarios by non-degenerate stars, the observed
correlation of host galaxy morphology with the rate of long GRBs
can be explained if long GRBs are produced in close binaries. If
stellar-wind model 3 is assumed for strong stellar winds in giant
spiral galaxies, this becomes true automatically, since the
minimum orbital period of a binary in which collapse of a
Wolf–Rayet star can occur is then about 3.5 day, condition
\ref{kerr} is not fulfilled, and all long GRBs occur in dwarf
irregular galaxies. If stellar-wind model 2 is adopted for strong
winds and model 1 for weak winds, the ratio of GRB rates depends
on the boundary between the dwarf and giant galaxies ({\it a}) and
the upper limit for the mass of the host galaxies of long GRBs
({\it b}). If we assume that the latter is approximately 10\% of
the mass of the Milky Way and $a\approx 5\cdot 10^9 M_{\odot}$ our
model for long GRBs fits the observational data well, with enough
leeway to allow for uncertainty in the critical orbital period for
producing long GRBs. If $b/a$ (for the wind combination we just
considered), a model in which long GRBs are products of the
evolution of the closest binaries cannot explain the observed
fractions of long GRBs in different types of host galaxies.

Thus, we conclude that the theoretical model for long GRBs
proposed in \cite{tutukov2003a,tutukov2004a}, fits the
observations within current uncertainties in the observational
data and theoretical models.

\end{document}